\documentclass[twocolumn,showpacs,preprintnumbers,amsmath,amssymb]{revtex4}

\usepackage{graphicx}

\usepackage{epsfig}

\begin{document}


\title{Asymptotic adaptive bipartite entanglement distillation protocol}

\author{Erik Hostens}
\email{erik.hostens@esat.kuleuven.be}
\affiliation{ESAT-SCD, K.U.Leuven, Kasteelpark Arenberg 10, B-3001 Leuven, Belgium}
\author{Jeroen Dehaene}
\affiliation{ESAT-SCD, K.U.Leuven, Kasteelpark Arenberg 10, B-3001 Leuven, Belgium}
\author{Bart De Moor}
\affiliation{ESAT-SCD, K.U.Leuven, Kasteelpark Arenberg 10, B-3001 Leuven, Belgium}
\date{\today}

\newcommand{\Z}{\mathbb{Z}}

\begin{abstract}
We present a new asymptotic bipartite entanglement distillation protocol that outperforms all existing asymptotic schemes. This protocol is based on the breeding protocol with the incorporation of two-way classical communication. Like breeding, the protocol starts with an infinite number of copies of a Bell-diagonal mixed state. Breeding can be carried out as successive stages of partial information extraction, yielding the same result: one bit of information is gained at the cost (measurement) of one pure Bell state pair (ebit). The basic principle of our protocol is at every stage to replace measurements on ebits by measurements on a finite number of copies, whenever there are two equiprobable outcomes. In that case, the entropy of the global state is reduced by more than one bit. Therefore, every such replacement results in an improvement of the protocol. We explain how our protocol is organized as to have as many replacements as possible. The yield is then calculated for Werner states.
\end{abstract}


\maketitle

\section{Introduction}
Quantum entanglement is an important resource in many applications of quantum cryptography and quantum communication. Some well-known examples are teleportation \cite{B3}, quantum key distribution \cite{ekert} and superdense coding \cite{B4}. These applications require pure and maximally entangled qubit pairs, called Bell state pairs, that are shared by two remote parties. One party prepares the Bell states and sends one qubit to the other party via some quantum channel. In a realistic setting, this channel is not perfect: uncontrollable influences of the environment (decoherence) will affect the qubit sent, resulting in qubit pairs that are in a mixed state and unsuitable for the application in mind.

Entanglement distillation is the process of applying local operations (local with respect to the parties) to the mixed state qubit pairs combined with classical communication (LOCC) in order to obtain pure Bell state pairs. Typically, we assume stationarity of the quantum channel, affecting all qubit pairs in the same way. As a result, we have $n$ copies of the same mixed two-qubit state $\rho$. Protocols like hashing or breeding \cite{Bennett,B2} have a net output of $m$ qubit pairs whose states approach pure Bell states if $n$ goes to infinity. We call such protocols \emph{asymptotic} and the fraction of distilled Bell states per initial copy the \emph{yield} $m/n$. Breeding differs from hashing by the use of an initial pool of predistilled Bell state pairs, but these protocols are known to be equivalent. The classical communication between the parties in both hashing and breeding is only in one direction. With two-way communication, higher yields can be achieved \cite{Bennett}. Indeed, the two parties can choose between alternative courses of the protocol based on information on intermediate stages. We call such a protocol \emph{adaptive}.

Entanglement distillation protocols, apart from being necessary for applications, are also interesting for theoretical purposes. The important entanglement measure \emph{entanglement of distillation} of $\rho$ is defined as the maximal asymptotic yield. It is lower bounded by the yields of all distillation protocols and in itself a lower bound for all sensible measures of entanglement \cite{hor,plenio}. Therefore, significantly improving distillation protocols brings us closer to a better understanding of the irreversible nature of entanglement manipulation.

Our protocol is based on the breeding protocol, with the incorporation of two-way communication. Until recently, the breeding or hashing protocol were the only existing asymptotic protocols, apart from the slightly better performing variant of Ref.~\cite{shor}. Adaptive upgrades of breeding/hashing mostly consist of breeding/hashing preceded by non-asymptotic recurrence-like schemes, resulting in higher yields only for low-fidelity states \cite{Bennett, DVD, man, mac}. Also the adaptive protocols of Ref.~\cite{ambainis} violate all kinds of one-way communication quantum error correction bounds, yet asymptotically do not perform any better than breeding/hashing. But Vollbrecht and Verstraete \cite{voll} came up with protocols that introduce two-way communication on an asymptotic level, improving breeding/hashing for all states. However, their protocols are rather ad hoc: further improvements are suggested by exhaustive searches over a rather untransparent decision space. We will explain the principles that are at the basis of the improvements and create new protocols that, by exploiting these ideas, outperform all existing schemes significantly.

Like all protocols mentioned, our protocols work for copies of a state $\rho$ that is diagonal in the Bell-basis, also called \emph{Bell-diagonal}. If $\rho$ is not Bell-diagonal, separate optimal single-copy distillation protocols can be applied to each copy to make them Bell-diagonal \cite{frank}. A nice feature of Bell-diagonal states is that they can be entirely interpreted in classical information theory. Indeed, the state $\rho^{\otimes n}$ is equivalent to a statistical ensemble of tensor products of Bell states. In the breeding protocol, information on $\rho^{\otimes n}$ is gathered from measurements on the Bell state pairs (ebits) of the initial pool, after letting them locally interact with $\rho^{\otimes n}$. One bit of information is gained for every ebit measurement, or equivalently, the entropy of $\rho^{\otimes n}$ is reduced by one bit. When the entropy of $\rho^{\otimes n}$ is reduced to zero, the ensemble has become a pure tensor product of Bell states. As will be explained in Sec.~\ref{secbreed}, breeding can be divided into successive stages of partial information extraction, yielding an equivalent protocol. The basic principle of our protocol is at every stage to replace measurements on ebits by measurements on a finite number of copies, whenever there are two equiprobable outcomes. It can be verified that the entropy of the global state is then reduced by more than one bit. This is because whenever an observable is measured, the state is projected onto the eigenspace of the observable, thereby eliminating the entropy associated with the outcomes of observables not commuting with the one measured. We will explain how our protocol is organized as to have as many replacements as possible. 

This paper is organized as follows. In the preliminary section~\ref{secprel}, an overview is given of the binary language in which our protocols are efficiently described. We also explain the two relevant ways of extracting information on an unknown tensor product of Bell states. In Sec.~\ref{secbreed}, we briefly explain the breeding protocol, partial breeding and the improvement of Ref.~\cite{voll}. In Sec.~\ref{secentred} we elaborate on the principle of \emph{entropy reduction}, on which our protocol is mainly based. The way equiprobable outcomes are forced and other ideas simplifying our protocols are then described in Sec.~\ref{secprotocol}. We also give a method for numerically calculating the yield. This is finally illustrated for Werner states in Sec.~\ref{secwerner}. We conclude in Sec.~\ref{secconclusion}.

\section{Preliminaries}\label{secprel}
In this section we give a short overview of the binary language in which distillation protocols are often expressed. For a detailed discussion and proofs of these results we refer to Refs.~\cite{D:03,DVD,Bennett,Hos}.
\subsection{Binary representation of Bell states, Pauli operators and Clifford operators}
Bell states can be represented by assigning two-bit vectors to the Bell states as follows
\begin{equation*}
\begin{array}{rcccl}
|\Phi^{+}\rangle &=& \frac{1}{\sqrt{2}}\left(|00\rangle+|11\rangle\right) &=& |B_{00}\rangle \\
|\Psi^{+}\rangle &=& \frac{1}{\sqrt{2}}\left(|01\rangle+|10\rangle\right) &=& |B_{01}\rangle \\
|\Phi^{-}\rangle &=& \frac{1}{\sqrt{2}}\left(|00\rangle-|11\rangle\right) &=& |B_{10}\rangle \\
|\Psi^{-}\rangle &=& \frac{1}{\sqrt{2}}\left(|01\rangle-|10\rangle\right) &=& |B_{11}\rangle.
\end{array}\end{equation*}
We consider all Bell states shared by two parties $A$ and $B$. In the following, all ``local'' operations are local with respect to the partition into $A$ and $B$.
In an analogous way, the Pauli matrices are identified with two-bit vectors:
\begin{equation*}\begin{array}{rcccl}
I_2 &=& \left[\begin{array}{rr}1&0\\0&1\end{array}\right] &=& \sigma_{00} \\
\sigma_x &=& \left[\begin{array}{rr}0&1\\1&0\end{array}\right] &=& \sigma_{01} \\
\sigma_z &=& \left[\begin{array}{rr}1&0\\0&-1\end{array}\right] &=& \sigma_{10} \\
\sigma_y &=& \left[\begin{array}{rr}0&-i\\i&0\end{array}\right] &=& \sigma_{11}.
\end{array}\end{equation*}
For notational convenience, we will often denote a binary vector by a string (e.g. 1010 means $[1~0~1~0]^T$).
A tensor product of $n$ Bell states can then be described by a $2n$-bit vector, e.g. $|B_{010011}\rangle=|B_{01}\rangle\otimes|B_{00}\rangle\otimes|B_{11}\rangle$. The same rule applies for a Kronecker product of Pauli matrices. The \emph{Pauli group} is defined to contain all Kronecker products of Pauli matrices with an additional complex phase factor in $\{1,i,-1,-i\}$, called Pauli operators. In the following, we will only consider Hermitian Pauli operators and neglect overall phase factors.

For all $a,b,s,t\in\mathbb{Z}_2^{2n}$, the following relations hold:
\begin{eqnarray*}
\sigma_a\sigma_b &\sim& \sigma_{a+b}, \\
(I_{2^n}\otimes\sigma_t)|B_s\rangle &\sim& |B_{s+t}\rangle,
\end{eqnarray*}
where ``$\sim$'' denotes equality up to an overall phase factor \cite{D:03,DVD}. All addition of binary objects is done modulo 2. Two Pauli operators $\sigma_a$ and $\sigma_b$ commute if the \emph{symplectic inner product} $a^TPb$ is equal to zero, or 
\begin{equation*}
\sigma_a\sigma_b=(-1)^{a^TPb}\sigma_b\sigma_a, \quad\mbox{where}\quad P=I_n\otimes\left[\begin{array}{cc}0&1\\1&0\end{array}\right].
\end{equation*}

A Clifford operator $Q$ maps the Pauli group to itself under conjugation, and can be represented by a symplectic matrix $C\in\Z_2^{2n \times 2n}$:
\begin{equation*}
Q\sigma_a Q^{\dag}\sim\sigma_{Ca}.
\end{equation*}
Symplecticity of $C$ is expressed by $C^TPC=P$. In the context of distillation protocols, we have the following interesting result \cite{DVD}: let $Q$ be represented by $C$ and $Q^\ast$ be the complex conjugate of $Q$, then it holds:
\begin{equation}\label{LC}
(Q\otimes Q^\ast) |B_s\rangle \sim |B_{Cs}\rangle,~\mbox{for all}~s\in\Z_2^{2n}.
\end{equation}

\subsection{Information extraction}\label{secinfoex}
Information on an unknown tensor product of $n$ Bell states $|B_s\rangle,~s\in\Z_2^{2n}$, in the context of distillation protocols, is extracted under the form of an inner product $r^Ts$, where $r$ is an arbitrary nonzero $2n$-bit vector. We will call this action a \emph{parity check}. This can be done in two ways: 
\begin{enumerate}
\item by local Clifford operations on the tensor product and an appended ebit $|B_s\rangle\otimes |B_{00}\rangle$, followed by the local measurement of the ebit;
\item by directly performing local measurements on $|B_s\rangle$.
\end{enumerate}
We explain the two ways in more detail, and call them \emph{appended ebit measurement} (AEM) and \emph{bilateral Pauli measurement} (BPM) respectively.

By means of local Clifford operations (\ref{LC}), we first transform $|B_s\rangle\otimes |B_{00}\rangle$ into $|B_s\rangle\otimes |B_{0~r^Ts}\rangle$. The symplectic matrix $C$ that corresponds to this action is
\begin{equation*}
C=\left[\begin{array}{ccc|cc}  & & & & 0 \\ & I_{2n} & & Pr & \vdots \\ & & & & 0\\ \hline 0 & \cdots & 0 & 1 & 0 \\ & r^T & & 0 & 1 \end{array}\right].
\end{equation*}
Then, a $\sigma_z$ measurement is performed on both sides of the appended ebit. The product of the outcomes is equal to $(-1)^{r^Ts}$. Indeed, the outcomes of a $\sigma$ measurement performed locally on an ebit correlate as follows:
\begin{equation*}\begin{array}{c|ccc}
 & \sigma_x & \sigma_z & \sigma_y \\ \hline
|B_{00}\rangle & +1 & +1 & -1 \\
|B_{01}\rangle & +1 & -1 & +1 \\
|B_{10}\rangle & -1 & +1 & +1 \\
|B_{11}\rangle & -1 & -1 & -1
\end{array}\end{equation*}
It follows that the product of the outcomes of a bilateral (i.e. on both sides) measurement $\sigma_{Pr}$ on a tensor product of Bell states $|B_{s}\rangle$ equals 
\begin{equation*}
(-1)^{r^Ts+r^TUr}, \quad\mbox{where}\quad U=I_n\otimes\left[\begin{array}{cc}0&1\\0&0\end{array}\right].
\end{equation*}

An AEM does not affect the state $|B_s\rangle$. Therefore, this procedure can be repeated consecutively for different $r$, like in the breeding protocol. However, the same does not hold for a BPM. Because our protocol will consist of both methods in various combinations, we need to sort out how this can be done. In Ref.~\cite{Hos}, we showed that, in theory, a BPM is equivalent to the following procedure: 
\begin{enumerate}
\item perform local Clifford operations (\ref{LC}) that correspond to a symplectic $C$ of which the last row is $r^T$: such a $C$ can always be found, for every $r\neq 0$;
\item then, perform a bilateral $\sigma_z$ measurement on the last qubit pair;
\item finally, apply the inverse of the local Clifford operations of the first step.
\end{enumerate}
Note that the result is no longer a tensor product of Bell states, as the last of the qubit pairs is measured in the second step, leaving it in a separable state. Since an AEM leaves the state $|B_s\rangle$ unaffected, we only need to worry about the situation after a BPM. The only irreversible step applied is the measurement of the last qubit pair, which yields knowledge of $r^Ts$ but destroys any other information contained by this pair. After this step, we are left with the state $|B_{\bar{C}s}\rangle$, where $\bar{C}\in\Z_2^{2(n-1)\times 2n}$ is equal to $C$ without the last two rows. The only information on $|B_s\rangle$ left for us to extract is the information we can extract from $|B_{\bar{C}s}\rangle$. Clearly, we can perform parity checks yielding $a^T\bar{C}s$, for all $a\in\Z_2^{2(n-1)}$. This is equivalent to determining $q^Ts$, for all $q\in\Z_2^{2n}$ that satisfy $q^TPr=0$. Indeed, as $C$ is symplectic, all such $q$ are in the column space of $\left[\bar{C}^T~r\right]$, or $q=\bar{C}^Ta+\alpha r$, for some $a\in\Z_2^{2(n-1)}$ and $\alpha\in\Z_2$. Since $r^Ts$ was already determined, we know $q^Ts=a^T\bar{C}s+\alpha r^Ts$ by determining $a^T\bar{C}s$ from the new state. In general, every time we determine $r^Ts$ of $|B_s\rangle$ by a BPM, afterwards we can only access $q^Ts$ where $q^TPr=0$, whatever method we use. This should not come as a surprise, because when $q^TPr=1$, the Pauli measurements $\sigma_{Pr}$ and $\sigma_{Pq}$ anticommute, so their outcomes cannot be determined both.

In reality, after a BPM, we should continue working with the transformed state represented by $\bar{C}s$. But this requires knowledge of the whole matrix $C$, while the parity check is specified only by $r$. As explained in the previous paragraph, we can describe all future actions in terms of $s$: we only need to know which BPM have been done. This yields a much more transparent description of the procotol.

\section{Breeding improved}\label{secbreed}
In this section, we start by briefly explaining the breeding protocol, which was introduced in Ref.~\cite{B2}. Basicly, information on $n$ copies of a Bell-diagonal mixed state is extracted sacrificing ebits until the state is a pure tensor product of $n$ Bell states (i.e. zero entropy). We show then how the breeding protocol can be divided into successive stages of \emph{partial information extraction}, yielding an equivalent protocol. Depending on the outcome of one such stage, a different strategy can be applied, yielding a protocol that uses two-way communication. We call such a protocol \emph{adaptive}, as it adapts to intermediate outcomes. We will explain an improvement of the breeding protocol that has been found in this way by Vollbrecht and Verstraete~\cite{voll}. For details, we refer to Refs.~\cite{Bennett, B2, voll}.

The breeding protocol starts from $n$ copies of a Bell-diagonal mixed state
\[\rho = \sum\limits_{v\in\Z_2^2}p_v|B_{v}\rangle\langle B_v|.\]
The global state $\rho^{\otimes n}$ is equivalent to a statistical ensemble of pure states $|B_s\rangle,~s\in\Z_2^{2n}$, with corresponding probabilities $p_s$ (e.g. $p_{001101}=p_{00}p_{11}p_{01}$). Consequently, the state can be regarded as an unknown pure state $|B_s\rangle$. The goal now is to determine $s$. Once we have pinned down $|B_s\rangle$, we can transform the state to $|B_0\rangle$ by performing the unitary transformation $\sigma_s$ on the $B$ side. With probability approaching 1 for large $n$, this unknown $s$ is contained in the \emph{typical set} $\cal T$ that has $\approx 2^{nS(\rho)}$ elements \cite{CT}, where
\[S(\rho)=-\sum\limits_{v\in\Z_2^2}p_v\log_2 p_v.\]
Consecutive parity checks $r^Ts$, where all $r$ are random, each on average rule out half of $\cal T$. Consequently, to obtain zero entropy (i.e. only one candidate left), about $nS(\rho)$ AEM are needed, each at the cost of one ebit. Therefore, the yield of the protocol, which is the number of ebits that are distilled for every copy, is equal to $1-S(\rho)$.

Partial information on $s$ is extracted by restricting to parity checks $r^Ts$, where $r$ is of the form
\[r=r'\otimes a,\]
$a$ is some fixed and finite $m$-bit vector ($m$ is even and divides $2n$) and random $r'\in\Z_2^{2n/m}$ take over the role of $r$. We will call this technique \emph{partial breeding}. Note that it is completely specified by $a$. Therefore we will denote it by PB $a$. We illustrate how partial breeding works with an example. Let $a=1010$, and divide $s$ into vectors of $m=4$ bits (i.e. $m/2=2$ pairs). Every such $m$-bit vector $g$ is either an element of $0^{(a)}$, if $a^Tg=0$, or of $1^{(a)}$, if $a^Tg=1$. For this example, we have
\[\begin{array}{rcl}
0^{(a)} &=& \{0000,0001,0100,0101,1010,1011,1110,1111\}, \\
1^{(a)} &=& \{0010,0011,0110,0111,1000,1001,1100,1101\}.
\end{array}\]
We have for instance
\[\begin{array}{rccccccc}
s= & 0010 & 1110 & 0110 & 0011 & 0001 & 1101 & 0100\\
\in & 1^{(a)} & 0^{(a)} & 1^{(a)} & 1^{(a)} & 0^{(a)} & 1^{(a)} & 0^{(a)}.
\end{array}\]
In the same way as for breeding, a typical set can be associated with the distribution of $0^{(a)}$ and $1^{(a)}$. This set has $\approx 2^{\frac{2n}{m}S^{(a)}(\rho)}$ elements, where
\[S^{(a)}(\rho) = -p_{0^{(a)}}\log_2 p_{0^{(a)}}-p_{1^{(a)}}\log_2 p_{1^{(a)}}. \]
Therefore, we need $\approx \frac{2n}{m}S^{(a)}(\rho)$ AEM to determine $a^Tg$ for all m-bit vectors $g$ constituting $s$, with probability close to 1. For this example, we have 
\[\begin{array}{rcl}
p_{0^{(a)}} &=& p_{0000}+p_{0001}+\ldots+p_{1111}, \\
p_{1^{(a)}} &=& p_{0010}+p_{0011}+\ldots+p_{1101}.
\end{array}\]

We have considered partial information extraction on a sequence of identically and independently distributed random variables over the set $\{00,01,10,11\}$. But the same idea can also be applied to the sets $0^{(a)}$ and $1^{(a)}$. Once we have carried out the previous PB step, we know for every 4-bit vector (2 pairs), whether it is in $0^{(a)}$ or $1^{(a)}$. If we bring all vectors in $0^{(a)}$ together, again we have i.i.d. random variables over $0^{(a)}$, and again we could perform partial breeding, this time for instance PB $b=0101$. Combining this with for instance PB $c=1000$ for $1^{(a)}$, we get to know for every 4 bits in which of the following sets they are:
\[\begin{array}{c}
S_1=0^{(a)}\cap 0^{(b)}=\{0000,0101,1010,1111\}, \\ S_2=0^{(a)}\cap 1^{(b)}=\{0001,0100,1011,1110\}, \\
S_3=1^{(a)}\cap 0^{(c)}=\{0010,0011,0110,0111\}, \\ S_4=1^{(a)}\cap 1^{(c)}=\{1000,1001,1100,1101\}.
\end{array}\]
It can be verified that the total number of AEM needed in the first and second PB step of this example is equal to
\[-\frac{n}{2}\left(p_{S_1}\log_2 p_{S_1}+\ldots+p_{S_4}\log_2 p_{S_4}\right),\]
which is exactly the entropy that is associated with the partition into $S_1,S_2,S_3,S_4$ times the number of 4-bit vectors in $s$. This is a consequence of the fact that \cite{CT}
\[S(A,B)=(-p_A\log_2 p_A-p_B\log_2 p_B)+p_A S(A)+p_B S(B).\]
So it is of no importance how a certain situation is attained, the number of AEM (= the cost in ebits) always equals the total information gain.
We can continue performing PB steps in this way until all sets considered are singletons. We then have determined $s$ completely, at the cost of $nS(\rho)$ ebits.

Of course, there is no point in dividing the breeding protocol in successive stages of partial breeding. In Ref.~\cite{voll}, $0^{(a)}$ pairs are further purified by breeding, but the $1^{(a)}$ pairs are treated differently: on the first pair of every $1^{(a)}$ state, a BPM $10$ is performed, yielding the parity $10$ of this pair. As the pair is measured, it is lost, but the measurement also provides information on the second pair. This one is in $\{10,11\}$ if the outcome was $+1$ and in $\{00,01\}$ if the outcome was $-1$. So in both cases, we end up with a rank two Bell-diagonal state, for which it has been proved that the breeding protocol is optimal \cite{rains}. The yield of this protocol is calculated in Ref.~\cite{voll}, and turns out to be greater than that of breeding. But the reason why this necessarily must be so, remains obscure. We will shed light to this issue in the next section.

\section{Entropy reduction}\label{secentred}
The reason why the protocol of Ref.~\cite{voll} outperforms the breeding protocol, lies in the difference between an AEM and a BPM. If a parity check is performed on a finite number $m/2$ of pairs, represented by an ensemble of vectors $g\in\Z_2^m$, the resulting state will have lower entropy by a BPM than by an AEM. Next to extracting information under the form of the parity, a BPM results in the mapping of different vectors to the same new vector, resulting in an extra entropy reduction.

To see this, we recall the procedure to carry out a BPM explained in Sec.~\ref{secinfoex}. If $a^Tg$ is the parity we would like to know, we first perform local Cliffords represented by a symplectic $C\in\Z_2^{m\times m}$ of which the last row is $a^T$, followed by a bilateral $\sigma_z$ measurement on the last pair. This results in a new state (with one pair less) represented by $\bar{C}g$. By the measurement, we learn $a^Tg$, but we also lose $b^Tg$, where $b$ is the second last row of $C$. This loss causes all $g$ with the same result $\bar{C}g$ and outcome $a^Tg$ to be mapped to the same vector $\bar{C}g$. Note that the outcomes should be equal as well, otherwise one of the two is ruled out. From the symplecticity of $C$, it follows that $g$ and $g+Pa$ are mapped together. Indeed, $\bar{C}Pa=0$ and $a^TPa=0$. Consequently, the new state is represented by the ensemble of vectors $\bar{C}g$, with probabilities $p_{g}+p_{g+Pa}$. This addition of probabilities results in the extra entropy reduction.

\begin{figure}
\includegraphics[width=0.45\textwidth]{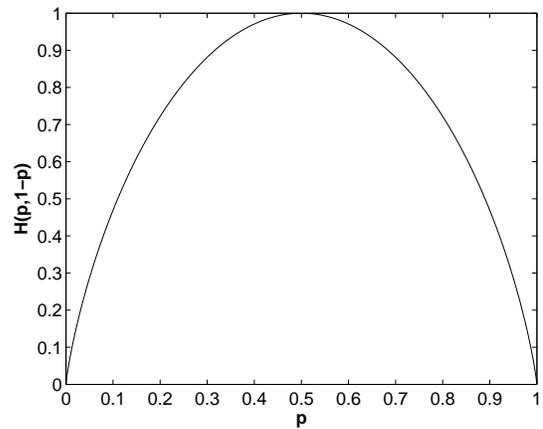}
\caption{\label{H} the binary entropy function $H(p,1-p)$.}
\end{figure}

Let us illustrate this with an example. We have two pairs represented by an ensemble of 4-bit vectors and we perform a BPM $1111$. We are left with only one pair represented by an ensemble of 2-bit vectors. The probabilities are
\[\begin{array}{cccc}
\frac{p_{0000}+p_{1111}}{p_{0^{(a)}}}, & \frac{p_{0011}+p_{1100}}{p_{0^{(a)}}}, & \frac{p_{0101}+p_{1010}}{p_{0^{(a)}}}, & \frac{p_{0110}+p_{1001}}{p_{0^{(a)}}}
\end{array}\]
if the outcome is $+1$, and
\[\begin{array}{cccc}
\frac{p_{0001}+p_{1110}}{p_{1^{(a)}}}, &  \frac{p_{0010}+p_{1101}}{p_{1^{(a)}}}, & \frac{p_{0100}+p_{1011}}{p_{1^{(a)}}}, & \frac{p_{0111}+p_{1000}}{p_{1^{(a)}}}
\end{array}\]
if the outcome is $-1$. Note that we do not identify these probabilities with the two-bit vectors $\bar{C}g$: all future actions are described entirely in terms of the original vectors $g$, as explained in Sec.~\ref{secinfoex}. If we would have used an AEM, then we would still have two pairs, but represented only by 8 vectors instead of 16, with probabilities
\[\begin{array}{cccccccc}
\frac{p_{0000}}{p_{0^{(a)}}}, & \frac{p_{1111}}{p_{0^{(a)}}}, & \frac{p_{0011}}{p_{0^{(a)}}}, & \frac{p_{1100}}{p_{0^{(a)}}}, & \frac{p_{0101}}{p_{0^{(a)}}}, & \frac{p_{1010}}{p_{0^{(a)}}}, & \frac{p_{0110}}{p_{0^{(a)}}}, & \frac{p_{1001}}{p_{0^{(a)}}}
\end{array}\]
if the outcome is $+1$, and
\[\begin{array}{cccccccc}
\frac{p_{0001}}{p_{1^{(a)}}}, & \frac{p_{1110}}{p_{1^{(a)}}}, &  \frac{p_{0010}}{p_{1^{(a)}}}, & \frac{p_{1101}}{p_{1^{(a)}}}, & \frac{p_{0100}}{p_{1^{(a)}}}, & \frac{p_{1011}}{p_{1^{(a)}}}, & \frac{p_{0111}}{p_{1^{(a)}}}, & \frac{p_{1000}}{p_{1^{(a)}}}
\end{array}\]
if the outcome is $-1$. The average difference in entropy is equal to
\[\begin{array}{cl}
& \lbrack -p_{0000}\log_2 p_{0000}-p_{1111}\log_2 p_{1111} - \ldots \\
& -p_{0111}\log_2 p_{0111}-p_{1000}\log_2 p_{1000}\rbrack \\
+ & \lbrack (p_{0000}+p_{1111})\log_2(p_{0000}+p_{1111}) + \ldots \\
& +(p_{0111}+p_{1000})\log_2(p_{0111}+p_{1000})\rbrack
\end{array}\]
and is always positive. Indeed, for all $x,y\geq 0$, we have:
\begin{equation}\label{Hred}\begin{array}{c}
[-x\log_2 x-y\log_2 y] + [(x+y)\log_2(x+y)] \\
=(x+y)H(\frac{x}{x+y},\frac{y}{x+y}), 
\end{array}\end{equation}
where 
\[H(p,1-p)=-p\log_2 p-(1-p)\log_2(1-p)\]
is the binary entropy function, plotted in Fig.~\ref{H}.

This plot shows that the entropy reduction, given by the right hand side of Eq.~(\ref{Hred}), is larger the more the colliding vectors $g$ and $g+Pa$ are equiprobable. If one probability relative to the other becomes small, the entropy reduction vanishes. That is the reason why the hashing protocol \cite{Bennett}, which is the same as breeding but the parity checks are BPM instead of AEM, has the same yield as the breeding protocol: again, we use the fact that almost all weight comes from vectors $s\in{\cal T}$. Since the $r$ are completely random, so are $s+Pr$. Therefore, the probabilities $\approx(p_{00}p_{01}p_{10}p_{11})^{n/4}$ of $s+Pr$ are infinitesimal (as $n$ is large) compared to the probabilities $\approx 2^{-nS(\rho)}$ of $s$ \cite{CT}. A variant of hashing \cite{shor}, where some of the BPM are on a finite number of copies resulting in a nonzero entropy reduction, performs slightly better than hashing.

It is clear that we should focus on BPM on small numbers of copies, because there lies the benefit of the entropy reduction. However, up till now, we have only spoken of the information gain, but we also have to take the cost into account. PB requires AEM, each at the cost of one ebit, whereas a BPM is at the cost of one of the copies. But as in the end all non-measured copies will be pure Bell states, this will not make the difference. By construction, every AEM in PB has equiprobable outcomes, and therefore yields one bit of information. The same does hold for a BPM if $r$ has infinite length and is random. Indeed, hashing is equivalent to breeding. But if we are to perform small non-random parity checks, the outcomes are not necessarily equiprobable and therefore yield less than one bit of information. If the outcomes are equiprobable, improvement is guaranteed. Note that the BPM $10$ on the first pair of two $1^{(a)}$ pairs does have equiprobable outcomes, which explains the improvement of Ref.~\cite{voll} over breeding. So in some way, we should try to spot as many finite equiprobable parity checks as possible and carry them out by BPM.

\section{Protocol}\label{secprotocol}
In the following, we will denote the all-zeros $m$-bit vector by $0_{m}$ and the all-ones vector by $1_{m}$. For any binary vector $g\in\Z_2^m$, we will denote $g+1_m$ by $\bar{g}$. Whenever a parity check $1_m$ has been performed on $m/2$ qubit pairs with outcome $\alpha\in\Z_2$ (we will call $\alpha$ the outcome instead of $(-1)^\alpha$), we will denote the resulting state by $[\alpha^{(m)}]$ if the parity check was a BPM and by $\alpha^{(m)}$ otherwise. Recall from Sec.~\ref{secentred} that the probabilities of $[\alpha^{(m)}]$ are (up to normalization) $p_g+p_{\bar{g}}$, whereas the probabilities of $\alpha^{(m)}$ are $p_g$, where in both cases all $g$ satisfy $1_m^Tg=\alpha$.

\subsection{Decoupling}
Learning the parity of a number of qubit pairs by partial breeding or BPM causes statistical dependence of the pairs involved, which makes the continuation of the protocol very complicated. However, this statistical dependence can be undone, which we refer to as \emph{decoupling}. The idea of decoupling is best explained by an example. Suppose by PB $1111$, we learn for every two copies of a Bell-diagonal qubit pair its state $\alpha^{(4)}$. Where the states of the copies were independent before, this obviously no longer holds afterwards. But if next we perform PB $11$ on all first pairs, yielding for a particular first pair its state $\beta^{(2)}$, where the state of both pairs was $\alpha^{(4)}$, we now have two independent pairs $\beta^{(2)}$ and $(\alpha+\beta)^{(2)}$. Indeed, we have learned the parities $1111\rightarrow\alpha$ and $1100\rightarrow\beta$, which is equivalent to knowing $1100\rightarrow\beta$ and $0011\rightarrow\alpha+\beta$, or $11$ for both pairs. So where the first PB coupled the ensembles of the two pairs, the second decoupled them again.

The same does hold for PB $1111\rightarrow\alpha$ followed by BPM $11\rightarrow\beta$ on the first pair. This is equivalent to BPM $11\rightarrow\beta$ on the first pair and PB $11\rightarrow\alpha+\beta$ on the second pair. And it can be verified that BPM $1111\rightarrow\alpha$ followed by BPM $11\rightarrow\beta$ on the first pair is equivalent to BPM $11\rightarrow\beta$ on the first pair and BPM $11\rightarrow\alpha+\beta$ on the second pair. This idea was also used in the adaptive stabilizer code formalism of Ref.~\cite{ambainis}.

However, this decoupling rule does not hold for BPM followed by PB. Once we have carried out a BPM on a number of qubit pairs, we have statistical dependence not only by the knowledge of the overall parity, but also by the mapping together of vectors as explained in Sec.~\ref{secentred}. It is this dependence that we denote by square brackets. Although the knowledge on the parities decouples by PB, this mapping does not. As an example, let BPM $1111$ followed by PB $1100$ have outcome 0 and 1 on two particular pairs respectively. The resulting state of the pairs is $[1^{(2)}1^{(2)}]$ and has probabilities:
\[\begin{array}{ccc}
\frac{p_{0101}+p_{1010}}{p_{1^{(2)}}^2} & \mbox{and} & \frac{p_{0110}+p_{1001}}{p_{1^{(2)}}^2}.
\end{array}\]
Therefore, once a BPM is carried out on a number of qubit pairs, we have to take it into account until it is later decoupled by a BPM on some of the qubit pairs.

We summarize all scenarios (the parity check on $\alpha^{(2m)}$ is always $1_m0_m$):
\begin{equation}\label{up1}\begin{array}{cccc}
 & \mbox{state} & \mbox{outcome} & \mbox{resulting state}\\ \hline 
\fbox{PB} & \alpha^{(2m)} & \rightarrow 0: & 0^{(m)}\alpha^{(m)} \\
 & & \rightarrow 1: & 1^{(m)}\bar{\alpha}^{(m)} \\ \hline
\fbox{BPM} & \alpha^{(2m)} & \rightarrow 0: & [0^{(m)}]\alpha^{(m)} \\
 & & \rightarrow 1: & [1^{(m)}]\bar{\alpha}^{(m)}
\end{array}\end{equation}
If the considered state was connected to others by previous BPM, like in $[x~\alpha^{(2m)}~y]$, the state transforms as follows:
\begin{equation}\label{up2}\begin{array}{cccc}
 & \mbox{state} & \mbox{outcome} & \mbox{resulting state}\\ \hline 
\fbox{PB} & [x~\alpha^{(2m)}~y] & \rightarrow 0: & [x~0^{(m)}\alpha^{(m)}~y] \\
 & & \rightarrow 1: & [x~1^{(m)}\bar{\alpha}^{(m)}~y] \\ \hline
\fbox{BPM} & [x~\alpha^{(2m)}~y] & \rightarrow 0: & [0^{(m)}][x~\alpha^{(m)}~y] \\
 & & \rightarrow 1: & [1^{(m)}][x~\bar{\alpha}^{(m)}~y]
\end{array}\end{equation}

Note that decoupling is nothing more than linearity of parity checks. Whenever we have performed a number of parity checks, these generate a space of parity checks. Any generating set of this space is equivalent to the original set of parity checks. E.g. $\{0101,1010\}$ is equivalent to $\{1010,1111\}$. We will use decoupling parity checks because they result in a transparent distillation protocol.

\subsection{Parity checks with equiprobable outcomes}\label{secequi}
In Sec.~\ref{secinfoex}, we showed that, once we have performed a BPM, we have to make sure that all following parity checks commute with it. There is a way in which this is automatically achieved. All vectors of the form $x\otimes 11$ commute (we could also have taken $01$ or $10$). Indeed, for all $2n$-bit vectors $x,y$, it holds:
\[\left(x\otimes\left[\begin{array}{c}1\\1\end{array}\right]\right)^T\left(I_n\otimes\left[\begin{array}{cc}0&1\\1&0\end{array}\right]\right)\left(y\otimes\left[\begin{array}{c}1\\1\end{array}\right]\right)=0.\]
Therefore, if we stick to parity checks of this form, we do not have to care about commutability any more. In this way, for every qubit pair we can find out whether it is $0^{(2)}$ or $1^{(2)}$. For now, let us assume we go up to this point but not further: we want to find an optimal way of reaching the point where every pair is determined as $0^{(2)}$ or $1^{(2)}$.

Whenever we spot parity checks with equiprobable outcomes, we should perform it by BPM. We will now explain how to do this. Suppose we have $m$ qubit pairs, determined as $1^{(2m)}$ by a previous parity check $1_{2m}$. Then the parity check $1_m0_m$ has equiprobable outcomes. Indeed, it holds that
\[1^{(2m)}=\begin{array}{l}
0^{(m)}1^{(m)}~\mbox{or} \\
1^{(m)}0^{(m)}.
\end{array}\]
Clearly, both possibilities have the same initial probability $p_{0^{(m)}}p_{1^{(m)}}$ or $1/2$ after normalization. Therefore, performing the parity check $1_m$ on the left half yields the parities of both halfs and this information equals one bit. By performing a BPM, we have the extra entropy reduction. Furthermore, this BPM decouples the two halves of the state.

However, if the $m$ pairs are
\[0^{(2m)}=\begin{array}{l}
0^{(m)}0^{(m)}~\mbox{or} \\
1^{(m)}1^{(m)},
\end{array}\]
we do not have equiprobable possibilities. With a little trick, we still are able to force an equiprobable outcome parity check. Two states of this kind can be written as
\[0^{(2m)}0^{(2m)}=\begin{array}{l}
0^{(m)}0^{(m)}~1^{(m)}1^{(m)}~\mbox{or} \\
1^{(m)}1^{(m)}~0^{(m)}0^{(m)}~\mbox{or} \\ \hline
0^{(m)}0^{(m)}~0^{(m)}0^{(m)}~\mbox{or} \\
1^{(m)}1^{(m)}~1^{(m)}1^{(m)}
\end{array}.\]
With an extra PB $0_m1_{2m}0_m$, we can distinguish the first two possibilities from the last two (as indicated by the line). If the outcome is 1, again we have two equiprobable possibilities $0^{(m)}0^{(m)}1^{(m)}1^{(m)}$ and $1^{(m)}1^{(m)}0^{(m)}0^{(m)}$, that are separated by a BPM $1_m$ on one of the four $m$-bit vectors. If the outcome is 0, the possibilities are not equiprobable, but again we can bring two of these results together, with possibilities
\[\begin{array}{l}
0^{(m)}0^{(m)}0^{(m)}0^{(m)}~1^{(m)}1^{(m)}1^{(m)}1^{(m)}~\mbox{or} \\
1^{(m)}1^{(m)}1^{(m)}1^{(m)}~0^{(m)}0^{(m)}0^{(m)}0^{(m)}~\mbox{or} \\ \hline
0^{(m)}0^{(m)}0^{(m)}0^{(m)}~0^{(m)}0^{(m)}0^{(m)}0^{(m)}~\mbox{or} \\
1^{(m)}1^{(m)}1^{(m)}1^{(m)}~1^{(m)}1^{(m)}1^{(m)}1^{(m)}
\end{array}\]
and performing PB $0_{3m}1_{2m}0_{3m}$ separating the possibilities as indicated by the line, and so forth. Clearly, this trick can be repeated endlessly.

We calculate the average fraction $\eta(0^{(2m)})$ of $0^{(2m)}$ on half of which a BPM $1_m$ is performed (note that $\eta(1^{(2m)})=1$). The procedure explained in the previous paragraph is recursive: at each step, we combine two random variables with two possible values $x$ and $y$ ($p_x+p_y=1$). The variables of the next step are $xx$ and $yy$, and so on. Therefore, it is possible to calculate $\eta(0^{(2m)})$ in a recursive way. Let $t$ be the probability to reach the situation under consideration and $k$ the total number of $0^{(2m)}$ involved in the present step. Initially, we have
\begin{eqnarray*}
t &=& 1 \\
p_x &=& \frac{p_{0^{(m)}}^2}{p_{0^{(m)}}^2+p_{1^{(m)}}^2} \\
p_y &=& \frac{p_{1^{(m)}}^2}{p_{0^{(m)}}^2+p_{1^{(m)}}^2} \\
k &=& 2.
\end{eqnarray*}
From the procedure explained in the previous paragraph, we have the following recursion relation:
\begin{eqnarray*}
t &\leftarrow& t(p_x^2+p_y^2) \\
p_x &\leftarrow& \frac{p_x^2}{p_x^2+p_y^2} \\
p_y &\leftarrow& \frac{p_y^2}{p_x^2+p_y^2} \\
k &\leftarrow& 2k.
\end{eqnarray*}
At each step, we have a probability $2p_xp_y$ that one of the $m$-bit vectors involved is detemined by BPM. So each step yields another fraction $2tp_xp_y/k$ of $0^{(2m)}$ on half of which a BPM is performed. It can be verified that the total sum of these fractions over all steps is equal to
\begin{equation}\label{f}\begin{array}{c}
\eta(0^{(2m)}) = \sum\limits_{i=0}^{\infty}\frac{(vw)^{2^i}}{2^i\prod\limits_{j=0}^{i}(v^{2^j}+w^{2^j})} \\
\mbox{where}~v=\frac{p_{0^{(m)}}^2}{p_{0^{(m)}}^2+p_{1^{(m)}}^2}~\mbox{and}~w=\frac{p_{1^{(m)}}^2}{p_{0^{(m)}}^2+p_{1^{(m)}}^2}.
\end{array}\end{equation}
In practice, it suffices to truncate the procedure after a few steps, since the terms in the summation of Eq.~(\ref{f}) decrease exponentially fast.

\subsection{Numerical calculation of the yield}\label{secnum}
The protocol starts with PB $1_{2^{q+1}}$. The next step is an iteration of the procedure explained in Sec.~\ref{secequi}, for $m=2^{q},2^{q-1},\ldots 2$, where we use the update rules (\ref{up1}) and (\ref{up2}). For now, we will treat all $0^{(2m)}$ in the same way, i.e. we do not favour particular states being parity checked by BPM. As a consequence, every $0^{(2m)}$ has the same probability $\eta(0^{(2m)})$ of undergoing a BPM $1_m0_m$. We find that, from one step to the next, the states transform as follows:
\begin{equation}\label{update}\begin{array}{cccc}
\mbox{state} & & \mbox{transforms to} & \mbox{with probability} \\ \hline
0^{(2m)} & \rightarrow & [0^{(m)}]0^{(m)} & \eta(0^{(2m)})/2 \\
 & \rightarrow & [1^{(m)}]1^{(m)} & \eta(0^{(2m)})/2 \\
 & \rightarrow & 0^{(m)}0^{(m)} & \frac{p_{0^{(m)}}^2}{p_{0^{(m)}}^2+p_{1^{(m)}}^2}-\frac{\eta(0^{(2m)})}{2} \\
 & \rightarrow & 1^{(m)}1^{(m)} & \frac{p_{1^{(m)}}^2}{p_{0^{(m)}}^2+p_{1^{(m)}}^2}-\frac{\eta(0^{(2m)})}{2} \\ \hline
1^{(2m)} & \rightarrow & [0^{(m)}]1^{(m)} & 1/2 \\
 & \rightarrow & [1^{(m)}]0^{(m)} & 1/2
\end{array}\end{equation}
With these rules, we are able to calculate the frequencies (i.e. the expected number of occurrences per $2^q$ qubit pairs) of all possibilities from one step to the next. After the last step, we are left only with $0^{(2)}$ and $1^{(2)}$ pairs, in various combinations of BPM (denoted by square brackets). Within square brackets, permutations of pairs yield equivalent states. Therefore, we do not have to calculate the frequencies of all possibilities, but only up to a permutation of the pairs: between square brackets, only the number $n_0$ of $0^{(2m)}$ and $n_1$ of $1^{(2m)}$ matter. We denote this by $[n_0,n_1]$. The possibilities in the end are then:
\begin{equation}\label{poss}
\begin{array}{l}
0^{(2)}, 1^{(2)}, \\
\lbrack1,0\rbrack, \lbrack0,1\rbrack, \\
\lbrack2,0\rbrack, \lbrack1,1\rbrack, \lbrack0,2\rbrack, \\
\qquad\vdots \\
\lbrack2^{q},0\rbrack, \lbrack2^{q}-1,1\rbrack, \ldots, \lbrack0,2^{q}\rbrack,
\end{array}
\end{equation}
with frequencies $f(0^{(2)}),f(1^{(2)}),f([1,0]),\ldots,f([0,2^{q}])$. Note that these must satisfy
\[\sum\limits_{A}\Bigl(n_0(A)+n_1(A)\Bigr)f(A)=2^q,\]
where we define $n_0(0^{(2)})=1,~n_1(0^{(2)})=0$ and  $n_0(1^{(2)})=0,~n_1(1^{(2)})=1$. By partial breeding alone, $nS^{(2)}(\rho)$ ebits would have been sacrificed. Now, for every BPM, we have one ebit less that has been measured. Therefore, the total cost of ebits per qubit pair up to this point equals
\begin{equation}\label{ebits1}
S^{(2)}(\rho) - \frac{1}{2^q}\sum\limits_{[n_0,n_1]} f([n_0,n_1]) .
\end{equation}

But the protocol is not finished yet. Breeding is optimal for the pairs that have never been involved in some BPM, as they are independent rank two Bell diagonal states \cite{rains}. We show that breeding is optimal for all pairs. Although equiprobable parity checks can still be found, they will no longer result in an entropy reduction if carried out by a BPM. Indeed, all further parity checks $a$ must be entirely built of $01$ and $10$, because for every pair we already know the parity $11$. Therefore, $Pa$ too is built of $01$ and $10$. Since every pair is either $0^{(2)}=\{00,11\}$ or $1^{(2)}=\{01,10\}$, the mapping of vectors vanishes: one of the two vectors mapped to the same new vector has already been ruled out by the parity checks, because $0^{(2)}+01=0^{(2)}+10=1^{(2)}$. Deprived of the benefit of entropy reduction by BPM, the best thing left is to gain one bit of information for every measurement. The number of ebits needed per qubit pair equals the entropy per pair
\begin{equation}\label{ebits2}
\frac{1}{2^q}\left(\sum\limits_{A}f(A)S(A)\right)
\end{equation}
left in the overall state. It can be verified that
\begin{eqnarray}
\nonumber S(0^{(2)}) &=& H(q_{00},q_{11}), \\
\label{entropies} S(1^{(2)}) &=& H(q_{01},q_{10}), \\
\nonumber S([n_0,n_1]) &=& -\frac{1}{2}\sum\limits_{i=0}^{n_0}\sum\limits_{j=0}^{n_1} {n_0 \choose i}{n_1 \choose j} P(i,j)\log_2 P(i,j),
\end{eqnarray}
\[\begin{array}[t]{cl}
\mbox{where} &
\begin{array}[t]{ccc}
q_{00}=\frac{p_{00}}{p_{00}+p_{11}}, & \qquad & q_{11}=\frac{p_{11}}{p_{00}+p_{11}}, \\
q_{01}=\frac{p_{01}}{p_{01}+p_{10}}, & \qquad & q_{10}=\frac{p_{10}}{p_{01}+p_{10}},
\end{array} \\
& P(i,j)=q_{00}^i q_{11}^{n_0-i} q_{01}^j q_{10}^{n_1-j}+q_{00}^{n_0-i} q_{11}^i q_{01}^{n_1-j} q_{10}^j.
\end{array}\]
Now all non-measured qubit pairs are pure ebits. The fraction of non-measured pairs equals
\begin{equation}\label{nonmeas}
1-\frac{1}{2^q}\sum\limits_{[n_0,n_1]}f([n_0,n_1]).
\end{equation}
If we substract the total number of measured ebits, which is the sum of (\ref{ebits1}) and (\ref{ebits2}), from this value (\ref{nonmeas}), we get the yield of the protocol:
\begin{equation}\label{yield}
1-S^{(2)}(\rho)-\frac{1}{2^q}\left(\sum\limits_{A}f(A)S(A)\right).
\end{equation}

\subsection{Favouring BPM on a small number of pairs}\label{secsmall}
It can be verified that the entropy reduction is larger for a BPM on a small number of pairs than on a large number of pairs. In the first version of our protocol, we did not make use of this, since all $0^{(2m)}$ were treated equally. So there is still room for improvement. As an example, consider the following situation:
\[[***][*****]\]
where all ``$*$'' are either $0^{(m)}$ or $1^{(m)}$, and a parity check $1_m$ on one of them (with equiprobable outcomes) determines them all. Then it is better to do a BPM on one of the first three, resulting in
\[[*][**][*****]\]
than on one of the last five, resulting in
\[[***][*][****].\]
Indeed, it can be verified that $S([***])-S([**])$ is larger than $S([*****])-S([****])$.

We show how to increase the number of BPM on small numbers of pairs. At each step, we have $0^{(2m)}$ and $1^{(2m)}$, distributed over all possibilities. We carry out BPM $1_m0_m$ on all $1^{(2m)}$, so there the situation remains the same. But the same cannot be done for all $0^{(2m)}$: there the ones that are linked by BPM (i.e. in square brackets) to a small number of pairs, should be taken first. Every $0^{(2m)}$ is part of some state $A$, where $n_0$ is nonzero. We now order all possibilities $[n_0,n_1]$ according to increasing $n_0+n_1$ and on a second level according to increasing $n_0$. So for example $[5,3]<[6,2]<[4,5]$. We favour small $n_0$ on a second level because all $1^{(2m)}$ will be certainly reduced, on average resulting in smaller $n_0$ and $n_1$ in the end. We also define that all $[n_0,n_1]<0^{(2m)}$. Probably better orderings can be found, but we do not want to complicate things further. We define
\[L(A)=\frac{\sum\limits_{B<A}n_0(B)f(B)}{p_{0^{(2m)}}2^q/m}\]
and $U(A)$ with the same formula but ``$<$'' replaced by ``$\leq$''. $L(A)$ and $U(A)$ are the fractions of all $0^{(2m)}$ that are part of some $B<A$ and $\leq A$ respectively. Note that $L([1,0])=0$ and $U(0^{(2m)})=1$. We combine the $0^{(2m)}$ for the procedure explained in Sec.~\ref{secequi} as follows: first we divide all $0^{(2m)}$ in two equally large sets (i.e. both sets contain $p_{0^{(2m)}}n/m$ elements): every $0^{(2m)}$ of the first set is part of some $A\leq$ that of every element of the second set. Now every $0^{(2m)}$ of the first set is combined with one of the second set and PB $0_m1_{2m}0_m$ is performed. Whenever the outcome is $1$ (the probability of which is calculated in the same way as in Sec.~\ref{secequi}), a BPM $1_m0_m$ is performed on the first $0^{(2m)}$. All $0^{(2m)}0^{(2m)}$ with outcome $0$ are again divided in two halves, according to the ordening of every first $0^{(2m)}$. By continuing in this way, the fraction $\eta(0^{(2m)}|A)$ of $0^{(2m)}$, part of some $A$, on which a BPM $1_m0_m$ is performed, can be calculated, and equals
\begin{equation}\label{fA}
\eta(0^{(2m)}|A)=\sum\limits_{i=1}^{u(A)-1}z_i+\sum\limits_{i=u(A)}^{l(A)}\frac{2^{-i}-L(A)}{U(A)-L(A)}z_i
\end{equation}
\[\begin{array}{crclcrcl}
\mbox{where} & l(A) &=& \lfloor -\log_2 L(A)\rfloor, & & v &=& \frac{p_{0^{(m)}}^2}{p_{0^{(m)}}^2+p_{1^{(m)}}^2}, \\
 & u(A) &=& \lceil -\log_2 U(A)\rceil, & & w &=& \frac{p_{1^{(m)}}^2}{p_{0^{(m)}}^2+p_{1^{(m)}}^2}, \\
 & z_i &=& \frac{2(vw)^{2^{(i-1)}}}{\prod\limits_{j=0}^{i-1}(v^{2^j}+w^{2^j})}. & & & &
\end{array}\]
As in Eq.~(\ref{f}), the terms in the second summation in Eq.~(\ref{fA}) decrease exponentially fast. Therefore, when $l(A)$ is large, the procedure may be truncated after a number of steps. In the update rules (\ref{update}), $\eta(0^{(2m)})$ must be replaced by $\eta(0^{(2m)}|A)$. Note that we have different update rules for different possibilities $A$. With this, we end up with the same possibilities (\ref{poss}) but with different frequencies $f(0^{(2)}),f(1^{(2)}),f([1,0]),\ldots,f([0,2^{q}])$. To calculate the yield, we still use Eqs.~(\ref{entropies}) and (\ref{yield}).

\section{Illustration with Werner states}\label{secwerner}
\begin{figure}[h]
\includegraphics[width=0.45\textwidth]{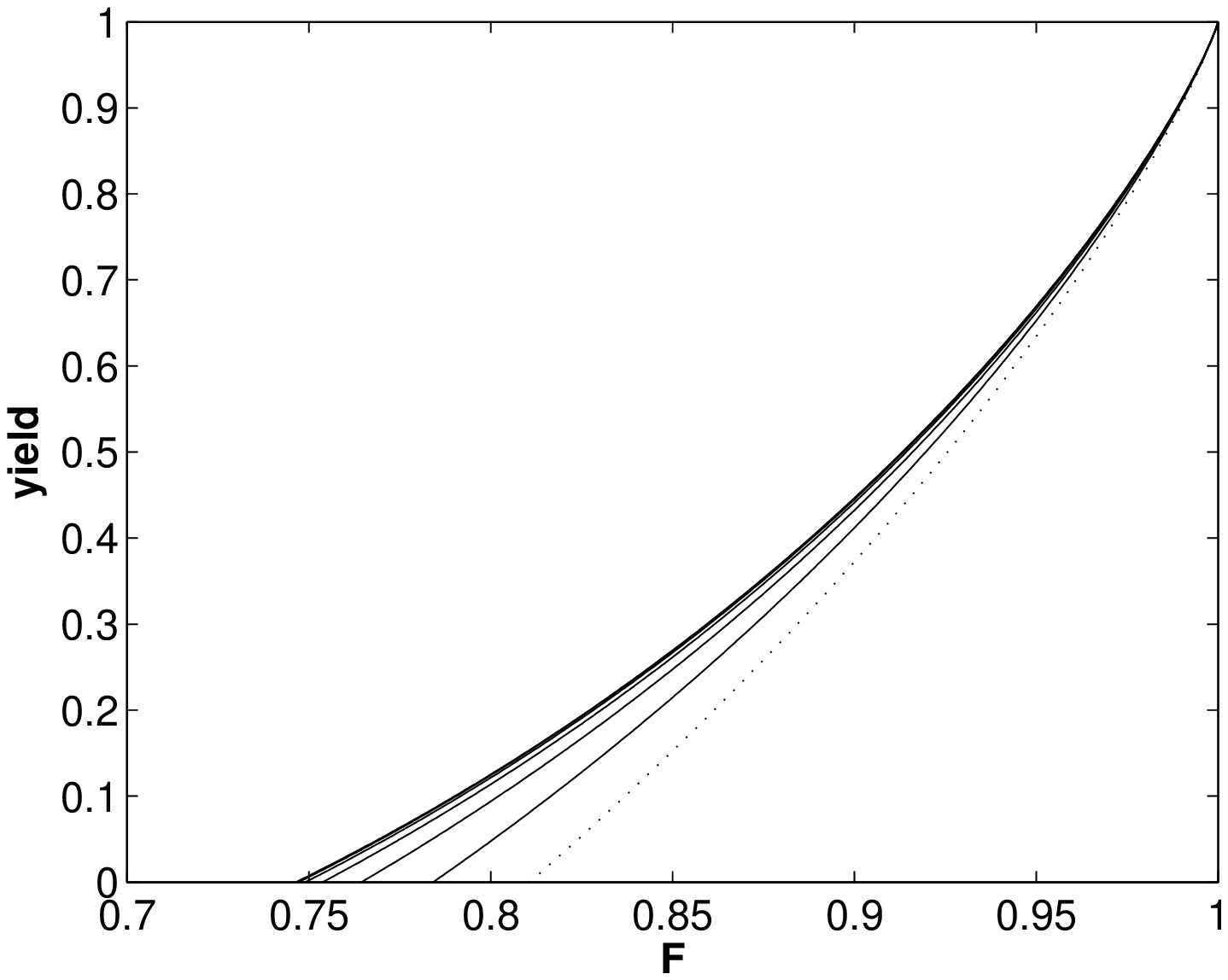}
\caption{\label{new1} the yields of the protocol of Sec.~\ref{secnum} (solid lines), for $q=1,2,3,4,5,6$, compared to the yield of breeding (dotted line). The yield increases with increasing $q$ and converges for large $q$.}
\includegraphics[width=0.45\textwidth]{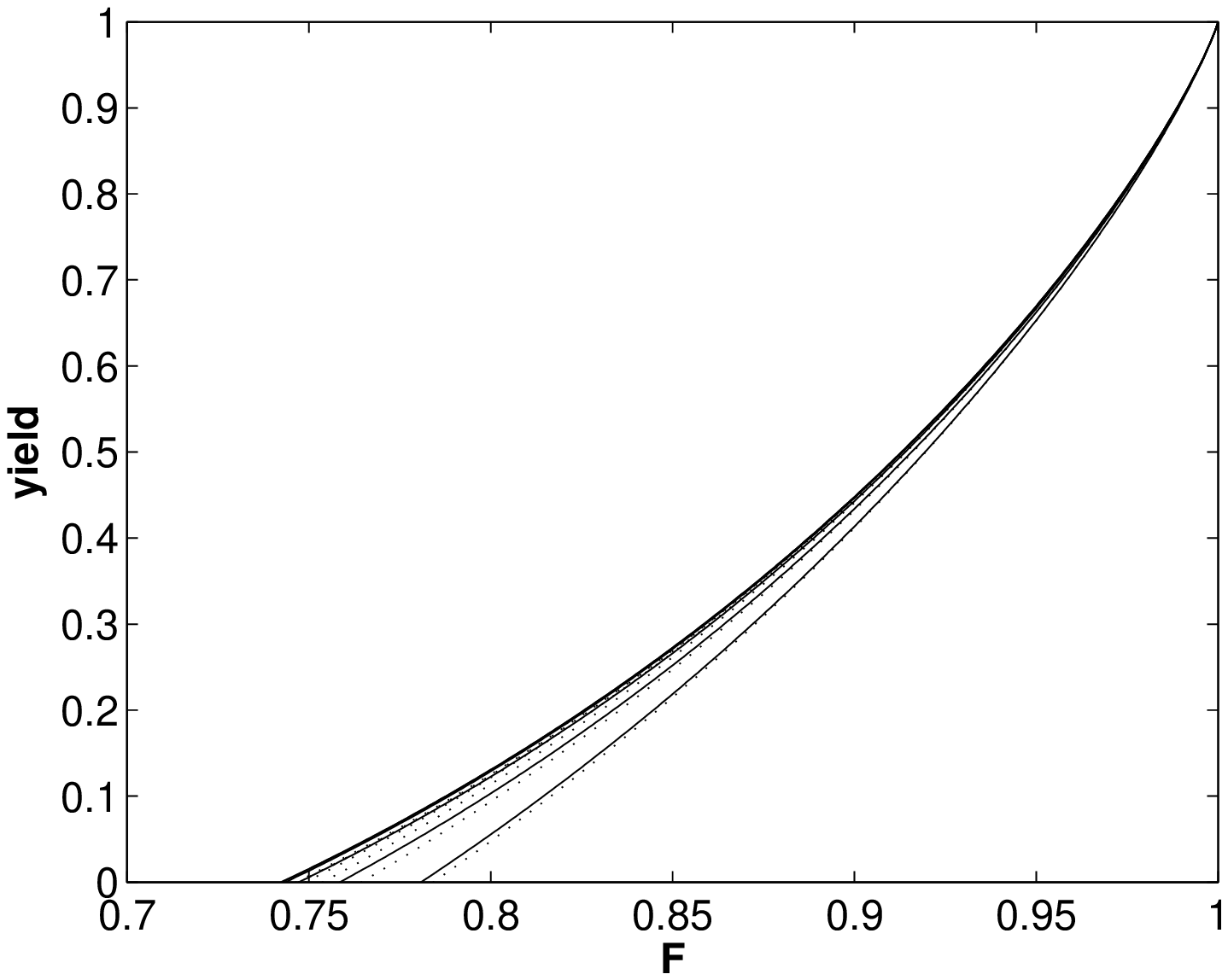}
\caption{\label{new2} the yields of the protocol of Sec.~\ref{secsmall}, where BPM on small numbers of pairs are favoured (solid lines), compared to the yields of that of Sec.~\ref{secnum} (dotted lines), for $q=1,2,3,4,5,6$. Again, the yield increases with increasing $q$ and converges for large $q$.}
\end{figure}

We have numerically calculated the yield of the protocols explained in Sec.~\ref{secprotocol} for Werner states. Werner states are Bell-diagonal states where $p_{00}=F$ and $p_{01}=p_{10}=p_{11}=\frac{1-F}{3}$. $F$ is also called the \emph{fidelity} of the state. Werner states are typically the result of one party preparing Bell states $|B_{00}\rangle$ and sending one qubit of the pair to the other party via the depolarization channel
\[\rho\mapsto F\rho+\frac{1-F}{3}\left(\sigma_x\rho\sigma_x^\dag+\sigma_y\rho\sigma_y^\dag+\sigma_z\rho\sigma_z^\dag\right).\]
In Figs.~\ref{new1} and \ref{new2}, we have plotted the yields of the protocols of Sec.~\ref{secnum} and \ref{secsmall}, for $q=1,2,3,4,5,6$. We truncate the procedure of Sec.~\ref{secequi} after 10 steps. We see that with increasing $q$, the yields of the protocols increase but converge. This is due to the fact that the entropy reduction is smaller for BPM on larger numbers of pairs. Also notice in Fig.~\ref{new2} that the yields of the protocol of Sec.~\ref{secsmall} are larger than the yields for corresponding $q$ of that of Sec.~\ref{secnum}.

We see that the yield of our best protocol is zero when $F\leq 0.7424$. This is better than breeding ($0.8107$), but in order to distill states with lower fidelity, we first have to apply a numer of iterations of recurrence \cite{Bennett}. Before every recurrence iteration, one-qubit local Clifford operations, yielding a permutation of the Bell states, are applied to each pair such that $p_{00}>p_{01},p_{10}\geq p_{11}$ for the transformed pairs \cite{Bennett,mac}. Recurrence itself consists of a BPM $1111$ on every two pairs, after which all remaining pairs where this parity check yielded 1, are discarded. The remaining pairs where the outcome was 0, have higher fidelity and are kept for a next iteration or for an asymptotic protocol. Note that the discarding can be interpreted as an extra BPM $1100$, which has equiprobable outcomes. 
\begin{figure}[h]
\includegraphics[width=0.45\textwidth]{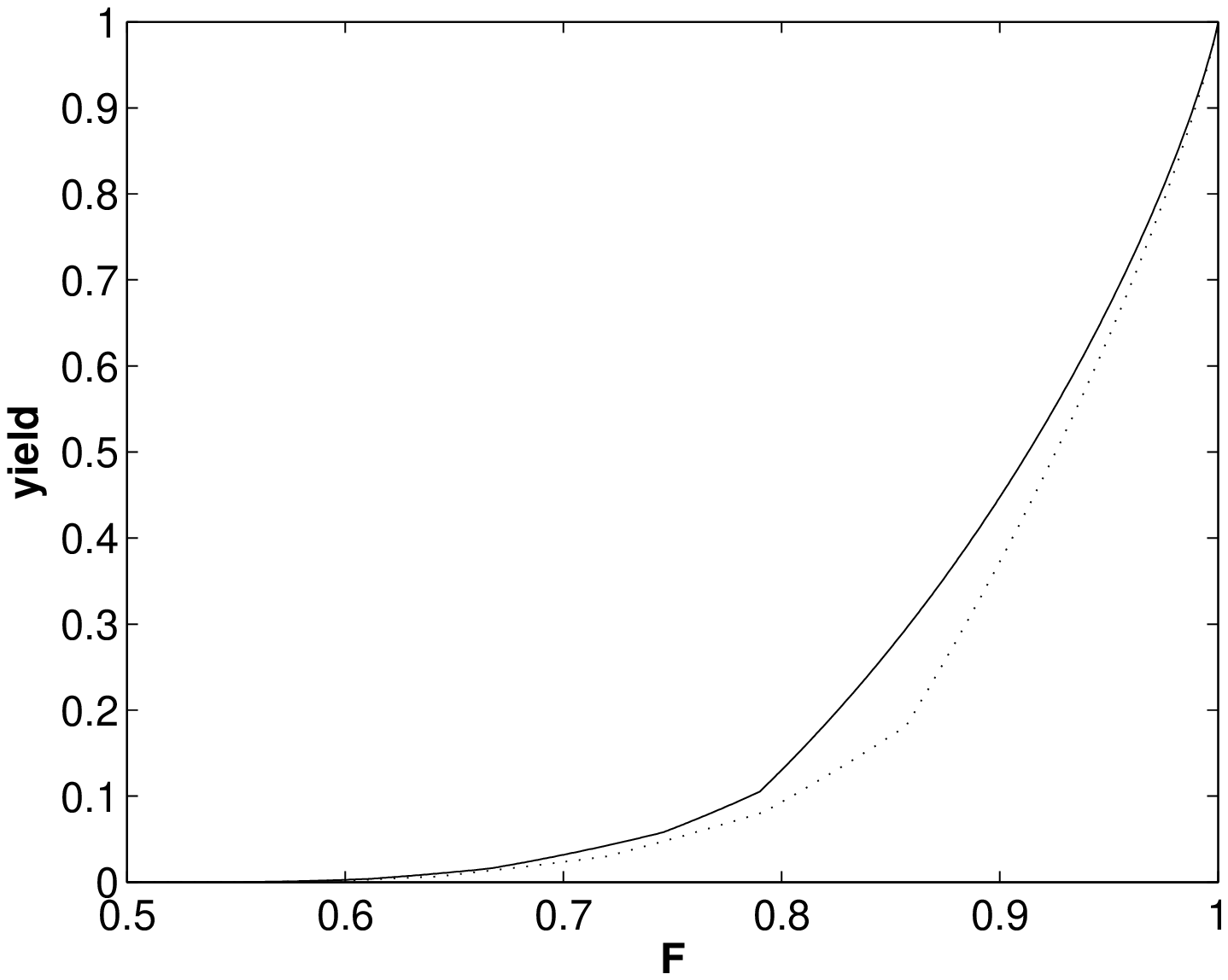}
\caption{\label{mac} the yield of our best protocol (solid line) and breeding (dotted line), both preceded by an optimal number of recurrence iterations.}
\includegraphics[width=0.45\textwidth]{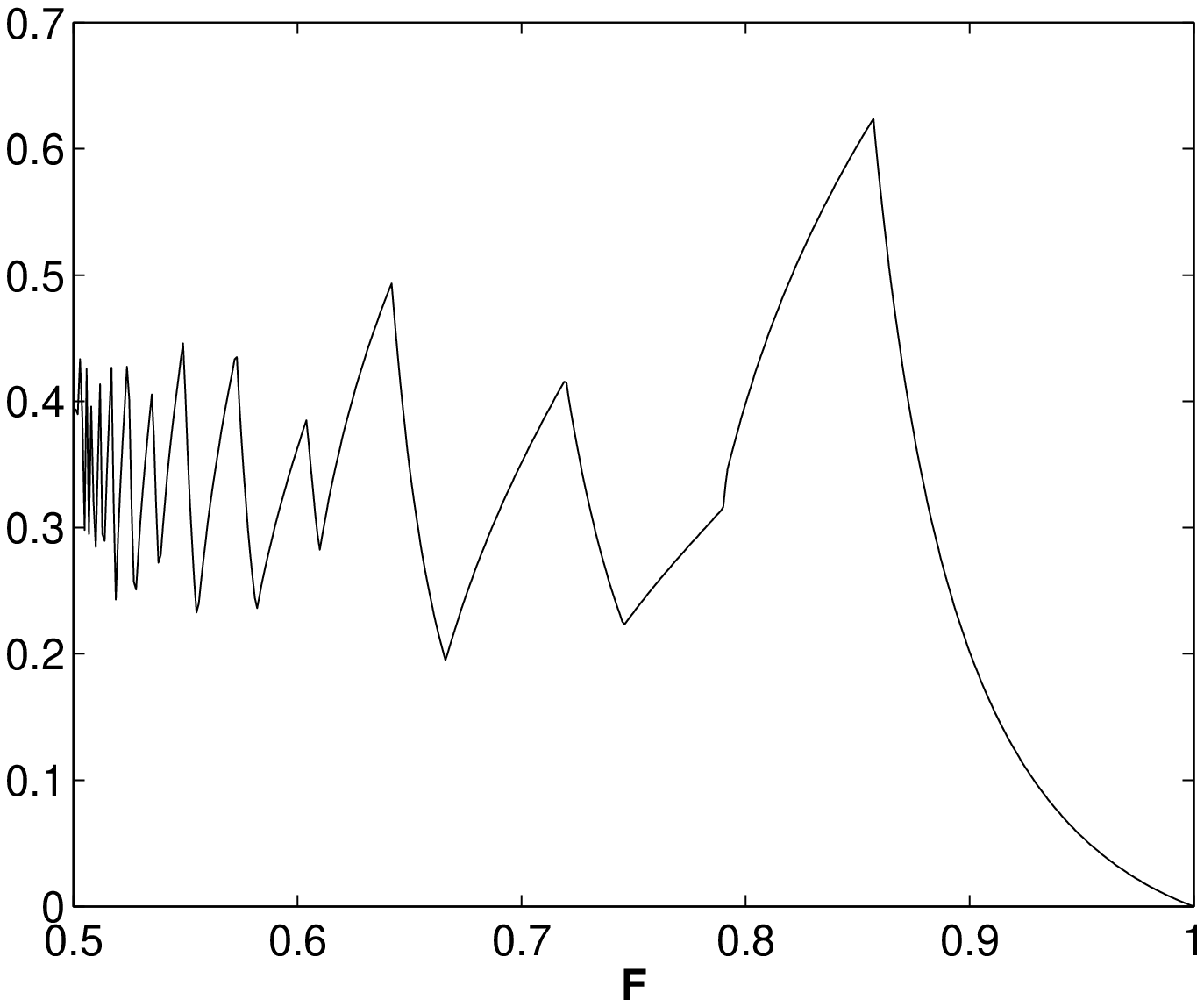}
\caption{\label{rel} the relative difference of the yields.}
\end{figure}
Therefore, the recurrence iterations before our protocol only improve it by the fact that also non-equiprobable parity checks are carried out by BPM. The not being maximal of the information gain is more than compensated by the entropy reduction for low-fidelity states. A next generation of protocols should incorporate a more complex criterion for BPM than merely equiprobable parity check outcomes, but we will not go deeper into that issue. We have compared the yield of breeding preceded by recurrence iterations to that of our protocol preceded by recurrence iterations in Fig.~\ref{mac}. The discontinuities in the slope are due to the fact that the optimal number of recurrence iterations is dependent on the fidelity. We have also plotted the relative difference in Fig.~\ref{rel}, which is the difference of the yields divided by the yield of breeding preceded by recurrence iterations. The sawtooth-like shape is caused by the fact that the discontinuities in the slopes of the yields do not coincide for the two protocols.

\section{Conclusion}\label{secconclusion}
We have presented a new asymptotic distillation protocol, that, based on the important principle of entropy reduction, outperforms all previous asymptotic protocols. Doing so, we have shed light on issues that were not clear before, such as the reason of the benefit of recurrence. Although we cannot claim to approach the entanglement of distillation, we certainly have tightened its lower bound. We also have mentioned roads that are still open for investigation. However, we feel that searching for further improvement will result in highly complicated protocols, possibly the product of an exhaustive search in a superexponential decision space.

\begin{acknowledgments}
Research funded by a Ph.D. grant of the Institute for the Promotion of Innovation through Science and Technology in Flanders (IWT-Vlaanderen).
BDM acknowledges the Katholieke Universiteit Leuven, Belgium. Research supported by Research Council KUL: GOA AMBioRICS, CoE EF/05/006 Optimization in Engineering, several PhD/postdoc \& fellow grants; Flemish Government: FWO: PhD/postdoc grants, projects, G.0407.02 (support vector machines), G.0197.02 (power islands), G.0141.03 (Identification and cryptography), G.0491.03 (control for intensive care glycemia), G.0120.03 (QIT), G.0452.04 (new quantum algorithms), G.0499.04 (Statistics), G.0211.05 (Nonlinear), G.0226.06 (cooperative systems and optimization), G.0321.06 (Tensors), G.0553.06 (VitamineD), research communities (ICCoS, ANMMM, MLDM); IWT: PhD Grants, GBOU (McKnow), Eureka-Flite2; Belgian Federal Science Policy Office: IUAP P5/22 (``Dynamical Systems and Control: Computation, Identification and Modelling'', 2002-2006); PODO-II (CP/40: TMS and Sustainability); EU: FP5-Quprodis; ERNSI; Contract Research/agreements: ISMC/IPCOS, Data4s, TML, Elia, LMS.
\end{acknowledgments}


\end{document}